# Crossing the light line

J.B. Pendry[1], P. A. Huidobro[2], M.G. Silveirinha[2], and E. Galiffi[1]

[1]The Blackett Laboratory, Department of Physics,
Imperial College London, London, SW7 2AZ UK

[2]Instituto de Telecomunicações, Instituto Superior Técnico-University of Lisbon,
Avenida Rovisco Pais 1, Lisboa, 1049-001 Portugal.

*Abstract*

We ask the question 'what happens to Bloch waves in gratings synthetically moving at near the speed of light?'. First we define a constant refractive index (CRI) model in which Bloch waves remain well defined as they break the light barrier, then show their dispersion rotating through 360 degrees from negative to positive and back again. Next we introduce the effective medium approximation (EMA) then refine it into a 4-wave model which proves to be highly accurate. Finally using the Bloch waves to expand a pulse of light we demonstrate sudden inflation of pulse amplitude combined with reversal of propagation direction as a luminal grating is turned on.

*Introduction*

Spurred on by experimental advances in rapid modulation both at optical [1,2], THz [3-5], and GHz [6,7,8] frequencies, in recent years interest has grown in systems where the parameters vary with time: in electromagnetic systems [9], the topic of the present paper, but also in acoustic systems [10,11,12]. In fact many of the concepts are quite general and apply to any wave motion whatever the system. One of the simplest static structures is the Bragg grating whose study has shown the way to photonic crystals and to a host of other electromagnetic devices. Translational symmetry permits the analysis of gratings in terms of Bloch waves and their dispersion is characterised by a Bloch wave vector which shows band gaps: ranges of frequency within which the wave vector is complex and where light is reflected from the structure. Therefore as an illustrative model we examine a synthetically moving Bragg grating of the form,

$$\varepsilon(x,t) = \varepsilon_1\left(1 + 2\alpha_\varepsilon \cos(gx - \Omega t)\right), \quad \mu(x,t) = \mu_1\left(1 + 2\alpha_\mu \cos(gx - \Omega t)\right) \quad (1)$$

This model has been much studied; earlier papers include [13-16]. It has been applied to study non-reciprocal systems [17] parametric amplification [18], topological aspects [19] and an extended discussion is to be had in [20]. More recent work of our own includes application to graphene [21], Fresnel drag [22], and novel amplification mechanisms [23].

The model has some remarkable properties when the grating velocity approaches the velocity of light, $\Omega/g \to c_1 = c_0/\sqrt{\varepsilon_1 \mu_1}$. Although the parameters have space-time translational symmetry and therefore can sustain Bloch wave solutions classified by a wave vector and frequency, $k, \omega$, in general this symmetry is broken when $\Omega/g \to c_1$ [24] and a phase transition into a localised state results.

In our earlier paper [24] we emphasised the importance of forward scattering in the localisation process. To isolate forward scattering as the mechanism responsible for the phase transition we eliminated back scattering by choosing $\alpha_\varepsilon, \alpha_\mu$ so that our model had constant impedance (CI) everywhere. This closed all the band gaps but left the phase transition intact. In this paper we do the opposite, keeping the refractive index constant (CRI) and modulating only the impedance,

$$\varepsilon(X) = \varepsilon_1\left(1 + 2\alpha_\varepsilon \cos(gX)\right), \quad \mu(X) = \mu_1/\left(1 + 2\alpha_\varepsilon \cos(gX)\right) \quad (2)$$

where $X = x - c_g t$ and the modulation velocity of the grating is $c_g = (1+\delta)c_0 / \sqrt{\varepsilon_1 \mu_1} = (1+\delta)c_1$. This choice eliminates the phase transition and preserves the validity of a Bloch wave description as $c_g$ transitions through the light line. We use $\delta$ to parametrise the grating velocity which passes the speed of light when $\delta = 0$.

In this paper we explore evolution of Bloch wave dispersion in near luminal gratings. Although devoid of a phase transition our model shows equally surprising consequences in this regime. We begin by using homogenisation theory to take a preliminary look at the dispersion relation.

*The homogenisation story*

Applying homogenisation theory of space-time media as reported in [25] we retrieve the following parameters. Working in a Galilean frame co-moving with the grating, $X = x - c_g t$,

$$\varepsilon_{mov} = \frac{1}{2\pi} \int_{gX=0}^{gX=2\pi} \frac{\varepsilon(X) d(gX)}{1-\varepsilon(X)\mu(X)c_g^2}, \quad \mu_{mov} = \frac{1}{2\pi} \int_{gX=0}^{gX=2\pi} \frac{\mu(X) d(gX)}{1-\varepsilon(X)\mu(X)c_g^2}$$

$$\xi_{mov} = -\frac{1}{2\pi} \int_{gX=0}^{gX=2\pi} \frac{\varepsilon(X)\mu(X)c_g d(gX)}{1-\varepsilon(X)\mu(X)c_g^2} \tag{3}$$

$$\varepsilon_{mov\|} = \left[\frac{1}{2\pi} \int_{gX=0}^{gX=2\pi} \frac{d(gX)}{\varepsilon(X)}\right]^{-1} \quad \mu_{mov\|} = \left[\frac{1}{2\pi} \int_{gX=0}^{gX=2\pi} \frac{d(gX)}{\mu(X)}\right]^{-1}$$

For the chosen modulation profile, (2), we can analytically solve these integrals and on transforming back to the laboratory frame we obtain the following effective parameters,

$$\varepsilon_{eff} = \frac{\varepsilon_1\left(1-c_g^2 c_1^{-2}\right)}{\sqrt{1-4\alpha_\varepsilon^2 - c_g^2 c_1^{-2}}} \sqrt{1-4\alpha_\varepsilon^2}, \quad \mu_{eff} = \frac{\mu_1\left(1-c_g^2 c_1^{-2}\right)}{\sqrt{1-4\alpha_\varepsilon^2 - c_g^2 c_1^{-2}}}$$

$$\xi_{eff} = -\frac{c_g c_1^{-2}\left(1-\sqrt{1-4\alpha_\varepsilon^2}\right)}{\sqrt{1-4\alpha_\varepsilon^2 - c_g^2 c_1^{-2}}} \tag{4}$$

$$\varepsilon_{eff\|} = \varepsilon_1 \sqrt{1-4\alpha_\varepsilon^2}, \quad \mu_{eff\|} = \mu_1$$

which give the following dispersion relationships for s and p polarisation in the effective medium approximation (EMA),

$$\varepsilon_{eff}\mu_{eff}\omega^2 = \frac{\mu_{eff}}{\mu_{eff\|}} k_y^2 + \left(k_s - \xi_{eff}\omega\right)^2$$

$$\varepsilon_{eff}\mu_{eff}\omega^2 = \frac{\varepsilon_{eff}}{\varepsilon_{eff\|}} k_y^2 + \left(k_p - \xi_{eff}\omega\right)^2 \tag{5}$$

The subscript "∥" refers to the component of the tensor parallel to the direction of motion, otherwise the component perpendicular. Also $k_s, k_p$ are the parallel components of the wave vector, $k_y$ the perpendicular component.

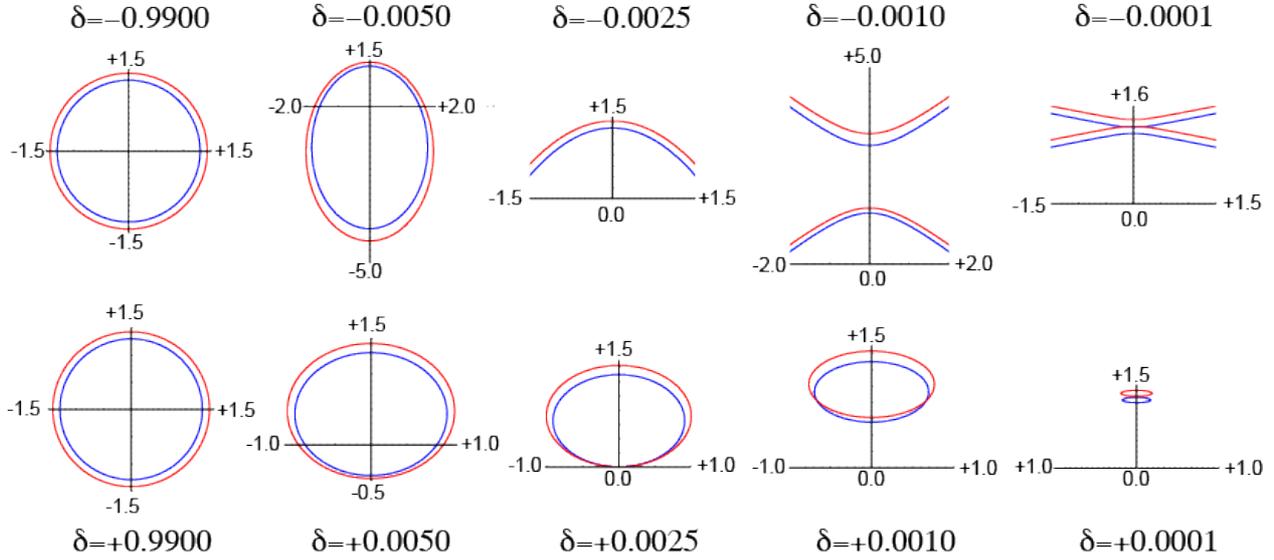

FIG. 1. Dispersion surfaces calculated for s-polarised radiation in the EMA for grating velocities close to the speed of light as indicated by $\delta$. We choose $\alpha_\varepsilon = 0.05$ and the surfaces are singular when $\delta \approx \pm\alpha^2 = \pm 0.0025$. $k_y$ lies along the abscissa and $k_s$, the component parallel to the direction of motion, along the ordinate. Both wave vectors are proportional to $\omega c_1^{-1}$ which we take to be unity. The top row approaches $c_1$ from below and the bottom row from above.

Fig. 1 shows dispersion of the wave vector for s-polarised radiation. Similar behaviour is shown by p-polarisation. The larger ellipse is calculated for $\omega' = 1.1 \times \omega$ and coloured red to show the velocity, given by the normal derivative of the dispersion surface. In a time independent system there are two Bloch waves with opposite velocities normal to the surface but this rule can be violated in time dependent systems as the figures show. In the following set of figures for the wave vectors $k_y$ lies along the abscissa and $k_s$ along the ordinate.

First consider the approach to $c_g = c_1$ from below. Starting with $c_g \approx 0$ at the top left of the figure we see a conventional isotropic dispersion: at each value of $k_y$ there are two Bloch waves headed with equal but opposite velocities carrying energy and information in opposite directions. As the grating speeds up an asymmetry develops. However notice that one of the two values of $k_s(k_y = 0)$ is little perturbed by the motion and this remains true for all of the figures: there is a Bloch wave at $k_s(k_y = 0) \approx 1.5$ in all of these figures. We call this a 'Spectator State' and its velocity lies in the same direction as the grating velocity. Further increase of the grating velocity elongates the ellipse until at the lower critical value of $\delta \approx -0.025$ the lower portion goes off scale to $-\infty$. Yet more velocity increase yields a hyperbolic dispersion surface the upper portion of which emerges at $+\infty$. We are now into strange territory since both Bloch waves have $k_s > 0$ and positive group velocities to match. This implies that pulses created from either or both of these waves can travel only in a positive direction which is a very strange state of affairs. We name these anomalous states 'Chimeras' after the monstrous creature of mythology. Further progression to $c_g = c_1$ results in a flattening of the hyperbola and a singularity at the crossing point where the hyperbola flips into two small ellipses on the far side.

Moving to the bottom row where $c_g > c_1$, $\delta > 0$ the ellipses grow from right to left, the lower values pushing towards $k_s = 0$ which is reached at the upper critical value of $\delta \approx +0.025$. The group velocity is infinite at this point and flips sign as $\delta$ increases beyond the critical value, returning to the conventional situation of two Bloch waves with opposed velocities and wave vectors. Finally for $c_g \gg c_1, \delta \ll 1$ we retrieve an isotropic dispersion surface and most anomalies disappear.

Let us explore the Chimera. Effective medium theory is exact for dispersion relations in the limit $\omega \to 0$, and when $k_y = 0$ we obtain the following analytic result for the velocity of the Chimera,

$$v_c = \frac{\omega}{k_s(k_y = 0)} = \frac{\alpha^2 + \delta}{\alpha^2 - \delta} c_1 \tag{6}$$

which shows the velocity rotating from a conventional $-c_1$ at low grating velocities through zero when $\delta = -\alpha^2$, staying positive until the next singularity when $\delta = +\alpha^2$, transitioning through $v_c = -\infty$ to negative velocities, and back to $-c_1$ at very high grating speeds. Notice that the branch cuts in $v_c$ which were a feature of our previous work [242], and which marked the zone of instability where Bloch waves are not defined, have now disappeared. As a consequence of our CRI assumptions, Bloch waves are well defined everywhere.

Although the EMA calculates the frequency dispersion exactly, problems arise when we turn to transmission and reflection coefficients. We can calculate these to first order in $\alpha_\varepsilon^2$ for a slab of grating thickness $d$,

$$t_{++} = e^{+ik_1 d}, \quad t_{-+} = \frac{\alpha_\varepsilon^2}{2}\left(e^{i(k_1 - k_2)d} - 1\right),$$
$$t_{--} = e^{-ik_2 d}, \quad t_{+-} = \frac{\alpha_\varepsilon^2}{2}\left(e^{i(k_1 - k_2)d} - 1\right) \tag{7}$$

where $t_{++}, t_{-+}$ are respectively the transmission and reflection coefficients for an incident wave travelling in a positive direction, and $t_{--}, t_{+-}$ the corresponding quantities for an incident wave travelling in a negative direction. Note that the EMA predicts that there is very little reflection in either case, and transmission is almost unity in both cases. Internal to the slab the Spectator state dominates forward transmission and the Chimera dominates backward transmission. In fig. 2 we compare these predictions with fully converged numerical transfer matrix calculations.

Fig. 2A shows that in contrast to the prediction of almost complete transparency the grating becomes highly reflecting for larger values of $d$. A clue to the reason is provided in fig. 2B where transfer matrix calculations show that the grating strongly excites waves of frequencies $\omega \pm \Omega$ and expels these waves from the grating.

*Crossing the light line*

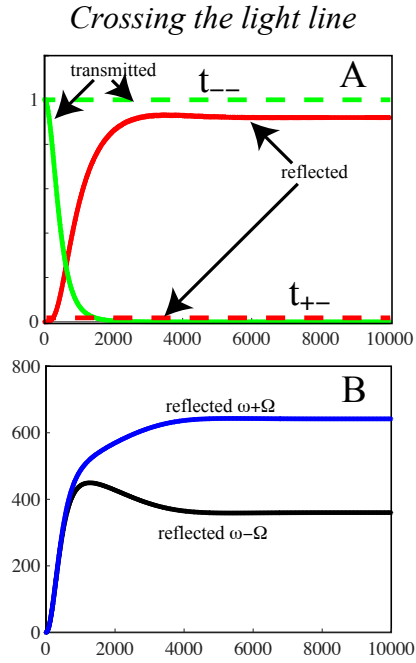

FIG. 2. Transmission and reflection of a wave travelling in a negative direction incident on a moving grating as a function of grating thickness, $d$. The grating travels at just less than the speed of light, $c_g = c_1(1+\delta)$ where $\delta = -0.002$. The incident frequency is chosen to be very low, $\omega = 0.0001$, and $g = 1.0, \alpha = 0.05$. Panel A compares transmission and reflection intensities at frequency $\omega$ as predicted by the EMA (dashed lines) compared to a fully converged transfer matrix calculation (full lines). Panel B shows the reflected intensities at $\omega \pm \Omega$ as given by the transfer matrix, The EMA makes no prediction for these intensities.

*Beyond EMA: the two Shepherds*

The EMA considers only two waves defined by their Bloch wave vectors. Moving to the next order we add two more forward travelling waves defined by the reciprocal lattice vectors of the grating. This 4-wave model comprises,

$$(+k,\omega) \quad (+k+g,\omega+\Omega) \quad (+k-g,\omega-\Omega) \quad (-k,\omega) \tag{8}$$

Fig. 3 shows the dispersion of these waves before hybridisation for a grating velocity slightly above the speed of light. Since the $\omega = -c_1 k$ wave has degeneracies with the two extra waves there will be strong hybridisation at these points which makes drastic changes to the three waves involved. The two extra waves we refer to as "Shepherds" for reasons which will become clear when we show the new dispersion diagrams.

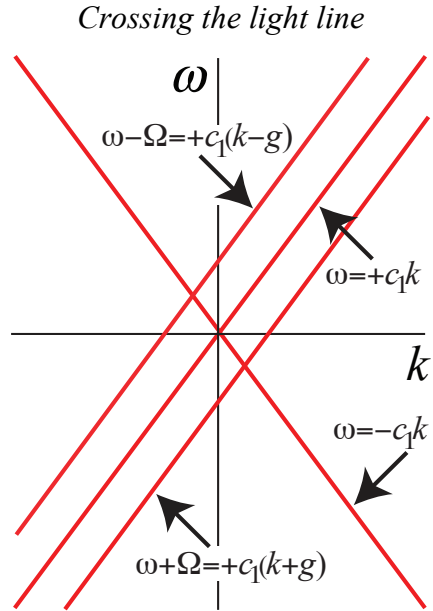

*Crossing the light line*

FIG. 3. Going beyond the EMA we add two further waves, the Shepherds, which are coupled by the grating.

The $\omega = +c_1 k$ state does not hybridise strongly with the two new waves and remains in the role of a Spectator just as the EMA implied.

As $c_g \rightarrow c_1$ the two Shepherds shown in fig. 3 crowd closer to the origin. Our specification of CRI means that there is no breakdown of Bloch symmetry and the Bloch wave vectors remain defined as $c_g$ transits through the velocity of light.

From Maxwell's equations we obtain an eigen-equation for the wave vector, $k$, given the frequency, $\omega$. We assume that $k_y = 0$ and truncate the matrix equation to contain the four waves specified in (7),

$$c_1^{-1} \begin{bmatrix} (\omega+2Bg) & +B\alpha g & 0 & +A\alpha g \\ -B\alpha g & \omega & +B\alpha g & 0 \\ 0 & -B\alpha g & (\omega-2Bg) & -A\alpha g \\ +B\alpha g & 0 & -B\alpha g & -\omega \end{bmatrix} \begin{bmatrix} b_{+1}^+ \\ b_0^+ \\ b_{-1}^+ \\ b_0^- \end{bmatrix} = k \begin{bmatrix} b_{+1}^+ \\ b_0^+ \\ b_{-1}^+ \\ b_0^- \end{bmatrix} \qquad (9)$$

where,

$$A = \frac{1}{2}(c_g + c_1), \quad B = \frac{1}{2}(c_g - c_1) \qquad (10)$$

and the eigenvectors contain the amplitudes of each of the four unhybridised waves.

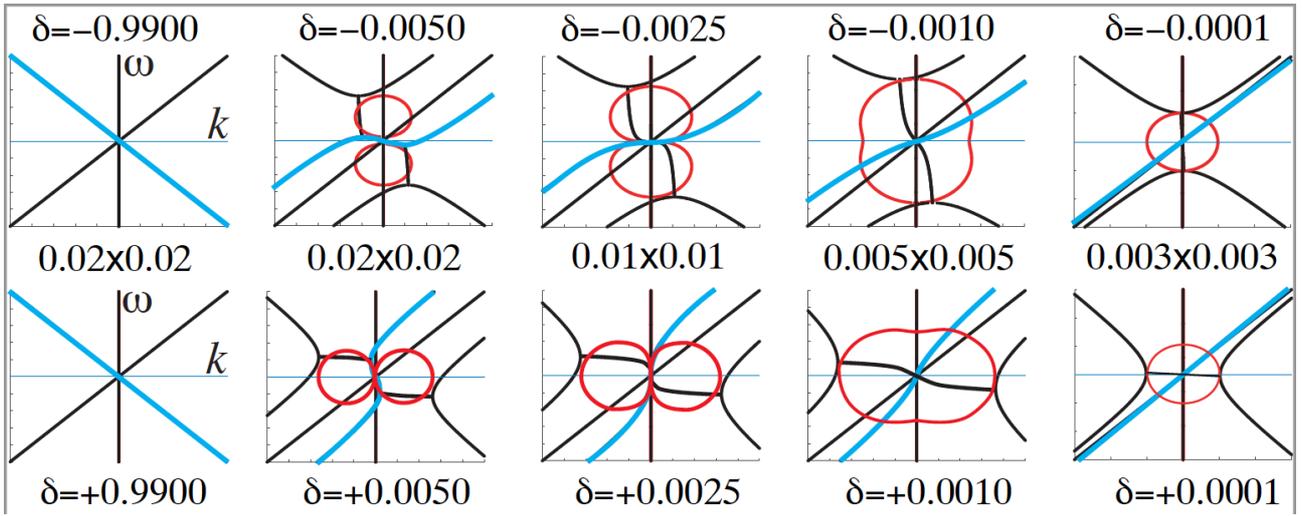

FIG. 4. Dispersion calculated for the four wave model for grating velocities close to the speed of light as indicated by $\delta$. As in fig. 1, we choose $\alpha_\varepsilon = 0.05$ and the curves are singular when $\delta \approx \pm\alpha^2 = \pm 0.0025$. Frequency, $\omega$, lies along the ordinate and the wave vector parallel to the direction of motion, $k_s$, along the abscissa. The top row approaches $c_1$ from below and the bottom row from above. The scale varies and is indicated in the central row of text which gives the total dimensions of the figures above and below the text. The Chimera is shown in cyan; red curves display the imaginary part of a wave vector in the upper set of figures, and the imaginary part of the frequency in the lower set.

In fig. 4 we show dispersion of the wave vectors parallel to the direction of motion calculated using the four wave model described above. Cyan colours the Chimera and red curves indicate the imaginary part of wave vectors, where present. Starting at the top left of the figure with a slowly moving grating we see two waves: a forward wave, and a backward wave that will morph into the Chimera. In the next figure in the top row the velocity is much closer to $c_1$ and the two Shepherd states are now in evidence, hybridising with the backward travelling Chimera to form two band gaps, but leaving dispersion of the forward travelling Spectator more or less untouched. At the critical value of $\delta = -0.0025$ a 3-fold degenerate critical point forms from the coalescence of the two imaginary wave vectors and the Chimera. The Chimera has zero group velocity at this point. Moving further to the right the degeneracy is lifted and the Chimera acquires a positive group velocity. When $c_g = c_1(1-0.0001)$ the Chimera is almost degenerate with the Spectator, and the band gap between the two Shepherds is near to closing which it does as the light line is crossed by $c_g$. This is another critical point where three bands are degenerate.

Now we move to the bottom of fig. 4 where $c_g > c_1$. Note the close relationship to the top row, but with $k$ and $\omega$ interchanged. Progressing to the left on the bottom row with increasing $c_g$ the group velocity of the Chimera near $\omega \approx 0$ increases. The two Shepherd states now have real wave vectors slightly offset from the abscissa so that when $\omega \approx 0$ all four waves have a positive group velocity. At the upper critical value of $\delta = +0.0025$ the Chimera's group velocity is infinite creating another tri-critical point where two wave vectors in the complex $\omega$ plane meet the Chimera. Thereafter the group velocity of the Chimera reverts to negative values with further increase of $\delta$. At high grating velocities the Shepherds have retreated far from the Chimera which has now been rotated by their influence through a full $2\pi$.

The four wave model is highly accurate and a fully converged matrix calculation with many more waves only changes the values shown here by $<1\%$. The EMA model described in the last section gives exact values for the wave vectors in the limit $\omega \to 0$. However the four wave model reveals the mechanism at work in rotating the Chimera.

*The tri-critical point and inflation*

At the tricritical points, which occur when $\delta = \pm\alpha^2$, 3 bands coalesce and hence the corresponding eigenstates of (8) are parallel to one another. This has important consequences for evolution of energy density in time. We consider a pulse of unit amplitude propagating in a motionless grating chosen to have very small values of $k$ and the grating is set in motion at $t = 0$. Since turning on the motion destroys translational symmetry in time but preserves translational symmetry, the pulse is expanded in terms of Bloch frequencies sharing the initial value of the wave vector $k$.

The fact that the Bloch waves are almost degenerate results in extremely large amplitudes for the three Bloch waves which when combined at $t = 0$ reduce to unit amplitude, but thereafter rapidly de-phase to inflate the pulse amplitude over a short time scale dictated by the residual differences in their frequencies. After a period of initial inflation subsequent behaviour depends on the Bloch frequencies excited. In fig. 5 we show some examples for the following parameters:

$$\alpha = 0.05, \quad \delta = \pm 0.002, \quad c_1 = 1.0, \quad g = 1.0, \quad k = 0.0001 \tag{11}$$

The waves are shaped into a Gaussian pulse which arrives at $t = 0$ with profile,

$$\exp(-\beta x^2), \quad \beta = 0.1 \times k^2 \tag{12}$$

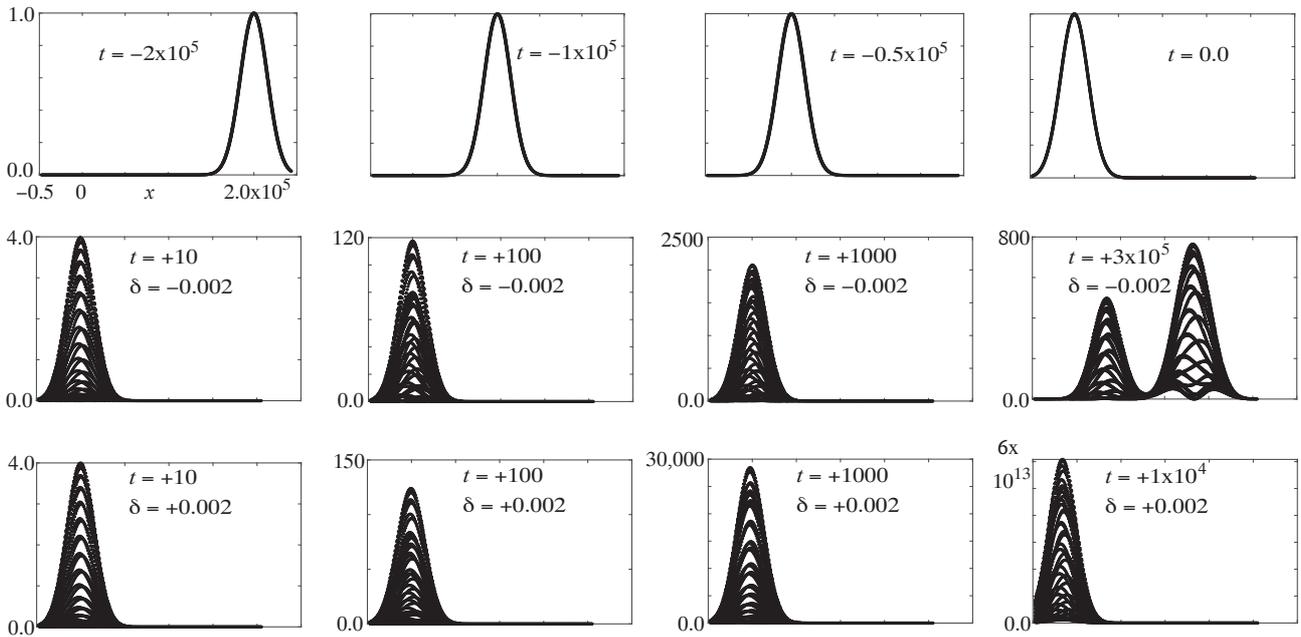

FIG. 5 Top row: intensity of a Gaussian pulse traveling in a backwards direction in a stationary grating as time advances from left to right (the *x* axis has the same scale in all graphs). The grating is set in motion at $t = 0$: in the second row with $\delta = -0.002$, $c_g < c_1$, and in the third row with $\delta = +0.002$, $c_g > c_1$. In each case the pulse excites nearly parallel Bloch eigenstates which rapidly de-phase giving rise to inflation of the pulse. For $\delta = -0.002$ the Bloch waves have real wave vectors and gradually separate. For $\delta = +0.002$ the wave vectors are complex: they stay put and grow exponentially in amplitude.

In the first row of fig. 5 the pulse is shown travelling with a negative velocity through a motionless grating. For the very small modulation, $\alpha = 0.05$, the motionless grating has almost no effect. As we move to the right along the row the pulse moves to the left until at $t = 0, x = 0$ the grating is set in motion and now the influence of the grating is very marked. We expand the pulse in terms of almost degenerate Bloch waves sharing the same value of $k$. Their frequencies can be found by interpolating fig. 4 between $-0.0025 < \delta < -0.001$.

Subsequent evolution of the pulse is shown in the second row for $\delta = -0.002$. The pulse comes to an immediate halt as no backwards travelling Bloch waves are available. Note the rapid inflation of the pulse amplitude shown by the increasing scale of the figures. For this value of $\delta = -0.002$ all of the relevant Bloch waves have real frequencies. Two have the same group velocity, somewhat greater than that of the third Bloch wave. The final figure in the second row shows the Bloch waves separating into two pulses each with a velocity in the reverse direction to the incident pulse. The time scale for inflation is relatively short and is related to the band gap between the Shepherd states, whereas the time to separate the components is much longer because the velocities concerned are all small as can be seen by examining fig. 4.

The multiple curves that appear in the second and third rows are an artifact of the plotting program: the Bloch waves have high spatial frequency components oscillating with a period $2\pi/g$, much more rapid than the sampling rate which is designed to reveal the profile rather than the detailed oscillations.

In the third row computed for $\delta = +0.002$ we see a similar initial picture with inflation dominating the scene. However two of the Bloch waves now have imaginary components to their frequencies and this causes them to grow exponentially with time as the final graph in fig. 5 shows.

*Conclusions*

We have introduced the CRI (constant refractive index) model in order to study evolution of Bloch wave dispersion as the grating velocity transitions through the light barrier. At least for weak modulation the model is well described by a 4 wave model: a largely undisturbed wave defined by $k, \omega$ travelling in the same direction as the grating and three further waves, $-k, \omega$ which we describe as a 'Chimera' because of its unusual behaviour, the other two, $-k + g, \omega + \Omega$; and $-k - g, \omega - \Omega$, which we name 'the Shepherds' because they interact strongly with the Chimera and are responsible for its extraordinary dispersion.

As $c_g = (1 + \delta)c_1 \to c_1$ two critical velocities are encountered. First at $\delta = -\alpha^2$ the Shepherds steer the Chimera group velocity to a positive value so that beyond this point all the states in our model have a positive group velocity until the upper threshold, $\delta = +\alpha^2$, is reached when the Chimera group velocity returns to a negative value. This raises interesting questions of how the system responds to an external stimulus.

At very low frequencies effective medium theory gives an exact description of the Chimera dispersion but is not successful in calculating the response to an externally incident wave for which the four wave model is required. The three eigenvectors, the Chimera and two Shepherds, exactly degenerate at the tri-critical points, are nearly parallel between the critical points and this gives rise to inflation: an incident pulse travelling through a stationary grating reverses its direction of travel and explodes when the grating starts to move.

This rich theoretical ecosystem exhibits a many surprising properties which we surmise may be realised in a variety of experimental situations.

*Acknowledgments*

P.A.H. and M.S. acknowledge funding from Fundação para a Ciencia e a Tecnologia and Instituto de Telecomunicações under Project UIDB/EEA/50008/2020. P.A.H. is supported by the CEEC

Individual program from Fundação para a Ciencia e a Tecnologia and Instituto de Telecomunicações with reference CEECIND/02947/2020. M.S. acknowledges support from the IET under the A F Harvey Prize and by the Simons Foundation under the award 733700 (Simons Collaboration in Mathematics and Physics, "Harnessing Universal Symmetry Concepts for Extreme Wave Phenomena"). E.G. acknowledges support from the Engineering and Physical Sciences Research Council through the Centre for Doctoral Training in Theory and Simulation of Materials (grant EP/L015579/1) and an EPSRC Doctoral Prize Fellowship (grant EP/T51780X/1). J.B.P. acknowledges funding from the Gordon and Betty Moore Foundation.

*References*

[1] Bruno, V., DeVault, C., Vezzoli, S., Kudyshev, Z., Huq, T., Mignuzzi, S. et al.
Negative Refraction in Time-Varying Strongly Coupled Plasmonic-Antenna-Epsilon-Near-Zero Systems.
*Phys Rev Lett*, **124**, 043902 (2020).

[3] Lira, H., Yu, Z., Fan, S., & Lipson, M.
Electrically driven nonreciprocity induced by interband photonic transition on a silicon chip
*Phys Rev Lett*, **109** (3), 033901 (2012).

[2] Anna C. Tasolamprou, Anastasios D. Koulouklidis, Christina Daskalaki, Charalampos P. Mavidis, George Kenanakis, George Deligeorgis, Zacharias Viskadourakis, Polina Kuzhir, Stelios Tzortzakis, Maria Kafesaki, Eleftherios N. Economou, and Costas M. Soukoulis
Experimental Demonstration of Ultrafast THz Modulation in a Graphene-Based Thin Film Absorber through Negative Photoinduced Conductivity
*ACS Photonics* **6**, 720-727 (2019).

[4] Christopher T. Phare, et al.
Graphene electro-optic modulator with 30 GHz bandwidth.
*Nature Photonics* **9,** 511-514 (2015).

[5] Wei Li, Bigeng Chen, Chao Meng, Wei Fang, Yao Xiao, Xiyuan Li, Zhifang Hu, Yingxin Xu, Limin Tong, Hongqing Wang, Weitao Liu, Jiming Bao, and Y. Ron Shen,
Ultrafast All-Optical Graphene Modulator,
*Nano Lett*., **14**, 2, 955–959, (2014).

[6] Planat, L., Ranadive, A., Dassonneville, R., Martínez, J. P., Léger, S., Naud, C. et al.
Photonic-Crystal Josephson Traveling-Wave Parametric Amplifier.
*Physical Review X*, **10**, 021021, (2020).

[7] A. M. Shaltout, V. M. Shalaev, and M. L. Brongersma,
Spatiotemporal light control with active metasurfaces,
*Science* **364**, eaat3100 (2019).

[8] V. Pacheco-Peña and N. Engheta,
Effective medium concept in temporal metamaterials,
*Nanophotonics* **9**, 379 (2020).

[9] Dani Torrent,
Strong spatial dispersion in time-modulated dielectric media,
*Phys. Rev.* **B102**, 214202, (2020).

[10] Choonlae Cho, Xinhua Wen, Namkyoo Park & Jensen Li,
Digitally virtualized atoms for acoustic metamaterials
*Nat. Comm.,* **11,** 1-8, (2020).

[11] D. Torrent, O. Poncelet, and J.-C. Batsale,
Nonreciprocal Thermal Material by Spatiotemporal Modulation,
*Phys. Rev. Lett.* **120**, 125501 (2018)

[12] M. Camacho, B. Edwards, and N. Engheta,
Achieving asymmetry and trapping in diffusion with spatiotemporal metamaterials,
*Nat. Commun.* **11**, 3733 (2020).

[13] E. Cassedy, A. Oliner,
Dispersion relations in time-space periodic media: Part (i) stable interactions.
*Proc. IEEE* **51**, 1342–1359 (1963).

[14] E. Cassedy,
Dispersion relations in time-space periodic media part (ii) unstable interactions.
*Proc. IEEE* **55**, 1154–1168 (1967).

[15] F. Biancalana, A. Amann, A.V. Uskov, E.P. O'Reilly,
Dynamics of light propagation in spatio-temporal dielectric structures.
*Phys. Rev.* **E75**, 046607 (2007).

[16] J. N. Winn, S. Fan, J. D. Joannopoulos, and E. P. Ippen,
Interband transitions in photonic crystals,
*Phys. Rev.* **B59**, 1551 (1999).

[17] D.L. Sounas, A. Alù,
Non-reciprocal photonics based on time modulation.
*Nat. Photonics* **11**, 774 (2017).

[18] T.T. Koutserimpas, A. Alù, R. Fleury,
Parametric amplification and bidirectional invisibility in PT-symmetric time-Floquet systems.
*Phys. Rev.* **A 97**, 013839 (2018).

[19] Lustig, E., Sharabi, Y., & Segev, M.
Topological aspects of photonic time crystals,
*Optica*, **5**, 1390-1395 (2018).

[20] Zoé-Lise Deck-Léger,, Nima Chamanara, Maksim Skorobogatiy, Mário G. Silveirinha and Christophe Caloz
Uniform-velocity spacetime crystals
*Advanced Photonics* 1 (5), 056002 (2019)

[21] E. Galiffi, P.A. Huidobro, J.B. Pendry,
Broadband nonreciprocal THz amplification in luminal graphene metasurfaces.
*Phys. Rev. Lett.* **123**, 206101 (2019).

[22] Paloma A. Huidobro, Emanuele Galiffi, Sebastien Guenneau, Richard V. Craster, and J. B. Pendry, Fresnel drag in space-time-modulated metamaterials,
Proc. Natl. Acad. Sci. **116,** 24943–24948 (2019).

[23] J. B. Pendry, P. A. Huidobro, and E. Galiffi,
Gain mechanism in time-dependent media
*Optica* **8** (5), 636-637 (2021).

[24] E. Galiffi, M.G. Silveirinha, P. A. Huidobro, and, J. B. Pendry,
Photon localization and Bloch symmetry breaking in luminal gratings
*Phys. Rev.* **B104** 014302 (2021).

[25] Paloma A. Huidobro, Mario G. Silveirinha, Emanuele Galiffi and J.B. Pendry,
Homogenisation theory of space-time metamaterials
*Phys. Rev. Appl.* **16,** 014044 (2021).